\documentstyle[12pt]{article}

\begin{document}

\def\lsim{\ \matrix{<\cr\noalign{\vskip-7pt}\sim\cr} \ }
\def\gsim{\ \matrix{>\cr\noalign{\vskip-7pt}\sim\cr} \ }

\begin{titlepage}

\hfill{PURD-TH-93-09}

\hfill{SISSA 87/93/A}

\hfill{DFPD/TH/93/42}

\vskip 1cm

\centerline{\Large \bf Baryogenesis through Collapsing String Loops }
\vskip 2mm
\centerline{\Large \bf in Gauged Baryon and Lepton Models}

\vskip 0.5cm

\centerline{{\bf Henry Lew$^{(a)}$}\footnote{email: lew\%purdd.hepnet@LBL.Gov}
{\bf and Antonio Riotto$^{(b)(c)}$}\footnote{email: riotto@tsmi19.sissa.it}}

\vskip 0.5cm

\noindent
\centerline{{\it (a) Physics Department, Purdue University, West Lafayette,
IN 47907-1396, U.S.A.}}

\vskip 2 mm
\noindent
\centerline{{\it (b) International School for Advanced Studies, SISSA,
via Beirut 2-4, I-34014 Trieste, Italy.}}

\vskip 2mm
\noindent
\centerline{{\it (c) Istituto Nazionale di Fisica Nucleare, Sezione di Padova,
I-35100 Padua, Italy.}}
\vskip 1cm

\centerline{\large\bf Abstract}
\vskip 1cm
\noindent
\baselineskip 20pt

A scenario for the generation of the baryon asymmetry in the early
Universe is proposed in which cosmic string loops, predicted by theories
where the baryon and/or lepton numbers are gauged symmetries, collapse
during the friction dominated period of string evolution. This provides
a mechanism for the departure from thermal equilibrium necessary to have a
nonvanishing baryon asymmetry. Examples of models are given where this idea
can be implemented. In particular, the model with the gauge symmetry
$SU(3)_{c}\otimes SU(2)_{L}\otimes U(1)_{Y}\otimes U(1)_{B} \otimes U(1)_{L}$
has the interesting feature where sphaleron processes do not violate the
baryon and lepton numbers so that no wash out of any initial baryon asymmetry
occurs at the electroweak scale.
\end{titlepage}
\leftline{\bf 1. Introduction}
\vskip 5mm
\baselineskip 20pt

The basic requirements for the generation of the baryon asymmetry
of the Universe (BAU) \cite{sak} are:
(1) baryon number violation, (2) $C$ and $CP$ violation and
(3) a departure from thermal equilibrium. In the past, models
of Grand Unified Theories (GUTs) in conjunction with the standard
hot big model of cosmology were found to have the above necessary
ingredients to explain the observed matter-antimatter asymmetry
of the Universe \cite{rev1}. However typical GUTs require relatively
high symmetry breaking scales (usually not too far away from the
Planck scale) making direct testing of such theories inaccessible
to currently known experimental techniques. It would be interesting
to find particle physics models which can realise the baryogenesis
scenario and can at the same time be tested experimentally in the
near future. Recent attempts to do this include baryogenesis at the
electroweak scale using just the standard model (SM) of particle physics
\cite{rev2}. In this paper we will explore an alternative mechanism
for generating the BAU and give examples of particle physics models
which can implement such a scenario.

This mechanism of generating the BAU is based on the collapse
of cosmic string loops. A variant of this mechanism has been
studied in Ref. \cite{bdh}. This will be described in section 2.
Then in the following sections we will give examples of particle
physics models which can realise this mechanism. The first example
will be a model with gauged baryon and lepton numbers. The crucial
aspect of this model is that since the baryon and lepton symmetries
are anomaly free then the sphaleron processes will not violate these
symmetries and hence any initial asymmetry produced will not be
washed out. This is in contrast with many of the GUT scenarios for
the generation of the BAU where the wash out of the initial asymmetry
is a common problem. The second example we will consider is a model
where the $(B-L)$ symmetry is gauged. The right-handed neutrinos in this
model are responsible for establishing an initial lepton number asymmetry.
The lepton number asymmetry is then converted to a baryon number ($B$)
asymmetry via the sphaleron. In both of these models the ultimate source
of the asymmetry comes from particle production as a result of the collapse
of cosmic string loops.

We conclude this section by commenting on the motivation for
the particular particle physics models we have chosen to discuss.
We feel that these models are interesting simply from the particle
physics point of view, let alone their implications for cosmology.
Firstly, these models deal with physics beyond the SM at scales well
below the GUT scale. They are simple in the sense that they rely only on
the concepts central to the theoretical framework of the SM. They
are (i) gauge symmetry and (ii) spontaneous symmetry breaking.
The success enjoyed by the SM in describing nature up to energies
hitherto explored by experiment has promoted these two concepts
to a status of fundamental importance. So much so, that in the case
of symmetries, any symmetry which is not gauged is viewed with
suspicion by some. For example, even though global symmetries
(continuous or discrete) appear in many particle physics models,
including the SM, they are considered to be less than fundamental
or at best ``accidental''. Since the origin of symmetries is still
an open question, such a theoretical prejudice might prove to be
wrong in the future. Nevertheless, in this paper we will pursue
this theoretical prejudice, perhaps to an extreme, by gauging all
or some combination of the known classical global symmetries of the SM.
\newpage
\leftline{\bf 2. Collapsing string loops mechanism}
\vskip 5mm

In this section we will describe a mechanism by which the
collapse of cosmic string loops can provide a source of
baryon or lepton number. The analysis will have some similarities
to that given in Ref. \cite{bdh} but because of the different
scales involved there will be some important differences.

Consider the Abelian Higgs model (which is the simplest example of a
model in which cosmic strings can form as a result spontaneous symmetry
breaking \cite{vilenkin}) where a complex Higgs field develops a vacuum
expectation value (VEV) $\langle H \rangle = \sigma e^{i\theta}$. The
arbitrary phase, $\theta=\theta(x)$, can vary in different regions of space.
For $H$ to be single valued $\theta$ must change by an integer multiple
of $2\pi$ around a closed loop. When the loop is shrunk to a point $\theta$
becomes undefined so that there exists a point where $\langle H \rangle =0$,
i.e., thin tubes of false vacuum get trapped somewhere inside the loop.
Such tubes of false vacuum or ``strings'' must either be closed or infinite
in length so that the closed loop cannot be contracted to a point without
encountering the tubes of false vacuum. Hence these strings are
topologically stable.

Once the cosmic strings have been produced
their evolution in time is demarcated by two periods. Initially the
strings will be moving in a dense medium so that their motion is
strongly damped and this is usually referred to as the friction
dominated period. After this time the friction becomes negligible
and the strings can move at relativistic speeds. The evolution of
strings is then mainly determined by the expansion of the Universe,
gravitational radiation of string loops and the intercommuting
of intersecting strings. The friction dominated period lasts from the
time of the phase transition, $t_c$, until the time
$t_* \simeq (G\mu)^{-1} t_c$,
where $G$ is Newton's constant and $\mu \simeq \sigma^2$ is the mass
per unit length of the string. In terms of temperatures,
$T_* \simeq{\sigma\over M_p}T_c$, where $M_p$ is the Planck mass and
$T_c \sim \sigma$ is the critical temperature of string formation.
For instance, if the phase transition takes place at scales less than
$\sigma \sim 10^{10}$ GeV then the friction dominated period will not
end before the temperature drops below about 10 GeV. Since in this paper
we are interested in relatively low symmetry breaking scales, only the
friction dominated period of string evolution will be relevant.
The situation depicted here is therefore different from that described
in Ref. \cite{bdh} where the breaking scale is very close to the GUT scale
and the friction dominated period lasts for a much shorter time. In such
a case the typical scale is the coherence length for a long string network
and the fact that its expansion is catching up with the Hubble radius must
be used.

Also note that since the gravitational effects of cosmic strings
depend on the dimensionless quantity $G\mu$, and $\sigma \ll M_p$,
then we expect that the usual cosmological effects associated with
strings (e.g. seeding for structure formation) will be negligible.

The basic idea for a nonzero $B$ and/or $L$ produced from strings is
that when strings collapse to a size comparable to its width the
microphysical forces of the string become important. As a result
the energy of the string will be released in the form of particle
production. We will assume for simplicity that all the energy
released from the collapsed string goes into the quanta of
the massive Higgs field. The energy of $N_Q$ quanta of mass
$M_h \simeq \lambda^{1\over 2}\sigma$ produced from the collapse of
a single string loop is $N_Q\lambda^{1\over 2}\sigma$, where
$\lambda$ is the quartic self-coupling of the Higgs field.
The energy of the string loop is given by $\mu\beta R$ where
$\beta \sim 2\pi$ and $R$ is the loop radius. Equating the two
expressions for the energy for $R \sim w$ results in
$N_Q \simeq \beta\lambda^{-1}$, where the thickness of the string
$w \simeq \lambda^{-{1\over 2}}\sigma^{-1}$ was used.

We will now proceed to estimate the number of loops contributing
to particle production. The size distribution of loops, for loops
of size $l$, is \cite{vach}
\begin{equation}
dn \sim \xi_c^{-4+\alpha} l^{-\alpha} dl,
\label{dist}
\end{equation}
where $\xi_c \simeq \left(\lambda\sigma\right)^{-1}$ is the
correlation length of the Higgs field and $\alpha$ was found
from Monte Carlo simulations of string systems to take on values
from $2$ to $2.5$ \cite{vach}.
Integrating Eq. (\ref{dist}) gives
\begin{equation}
n(t_w) \simeq {\xi_c^{-4+\alpha}\over (1-\alpha)}
\left\{ l(t_c)^{(1-\alpha)} - l(t_w)^{(1-\alpha)} \right\},
\label{ntw1}
\end{equation}
where $t_w$ is the time when $l(t_w) = w$ and $l(t_c)$ is
expected to be the largest sized loop that can collapse to
$l(t_w)$ in time.

To find the explicit time dependence of the loop size, consider
the motion of the string in the friction dominated epoch.
The typical damping time on the motion of the string is \cite{damp}
\begin{equation}
t_d \sim {\Delta P / l \over F / l}
\simeq {\mu\over n} \simeq {\sigma^2 \over k_n T^3},
\end{equation}
where the change in momentum per unit length of the string is
$\Delta P/l \sim \mu v$ and the force per unit length on the
string is $F/l \sim n v$. $v$ is the speed of the string relative
to the surrounding medium and $n = k_n T^3$ is the number density
with
\begin{equation}
k_n = {\zeta(3)\over \pi^2}
\sum_{i=bosons}g_i \left({T_i\over T}\right)^3
+{3\over 4}\sum_{i=fermions}g_i \left({T_i\over T}\right)^3,
\label{kn}
\end{equation}
where $\zeta(x)$ is the Riemann zeta function and $g_i$ is the
number of degrees of freedom of the particles in the dense medium.
Therefore the characteristic loop size at a time $t$ is
\begin{equation}
l(t) \sim t_d \simeq {\sigma^2\over k_n}
\left({t\over \kappa M_p}\right)^{3\over 2},
\label{lsize}
\end{equation}
where the time-temperature relation of $t = \kappa M_p T^{-2}$
for an expanding Universe in the radiation dominated epoch was
used. The rate at which the loops collapse can be estimated by
assuming
\begin{equation}
{dl\over dt} \sim -{dt_d \over dt}.
\end{equation}
This then gives
\begin{equation}
l(t_c) \simeq w + {\sigma^2\over k_n}\left\{
\left({t_w\over \kappa M_p}\right)^{3\over 2}
-\left({t_c\over \kappa M_p}\right)^{3\over 2} \right\},
\end{equation}
with $l(t_w) = w$. Substituting this into Eq. (\ref{ntw1})
results in
\begin{equation}
n(t_w) \simeq {\xi_c^{-4+\alpha}\over (1-\alpha)}
\left\{ \left[w + {\sigma^2\over k_n}\left\{
\left({t_w\over \kappa M_p}\right)^{3\over 2}
-\left({t_c\over \kappa M_p}\right)^{3\over 2} \right\}\right]
^{(1-\alpha)} - w^{(1-\alpha)} \right\}.
\label{ntw2}
\end{equation}
This is the number of loops per unit volume that will collapse to size $w$.

However, not all of these loops will contribute to generating
a nonzero $B$ or $L$ since the decays of the Higgs particles
will be compensated by their inverse decays at temperatures above
the freeze-out temperature, $T_F$, (corresponding to a time, $t_F$)
which is of the order of the Higgs boson mass. The relevant number density
of loops is then given by
\begin{equation}
n = \int_{t_F}^{t_{min}} f(t_w)
\left[{a(t_w)\over a(t_F)}\right]^3 dt_w,
\label{n}
\end{equation}
where
\begin{equation}
f(t_w) \equiv {dn(t_w)\over dt_w}
\end{equation}
and $a(t) \propto t^{1\over 2} \propto T^{-1}$ is the usual cosmic
scale factor in the radiation dominated epoch of the Universe. The
limits of integration correspond to their respective temperatures,
i.e.,
\begin{eqnarray}
t_{min} & \longleftrightarrow &
T_{min} \simeq k_n^{-{1\over 3}}\lambda^{1\over 6}\sigma
\nonumber \\
t_F & \longleftrightarrow & T_F \simeq \lambda^{1\over 2}\sigma,
\end{eqnarray}
$T_{min}$ is obtained by requiring $l(t) \geq w$
(see Eq. (\ref{lsize})). To simplify the evaluation of $n$,
notice that the ratio
\begin{equation}
\left[{a(t_{min}) \over a(t_F)}\right]^3
= \left({T_F\over T_{min}}\right)^3 \simeq k_n\lambda.
\end{equation}
This is expected to be of order one for ``typical'' values of the
parameters (e.g. $k_n \sim 10$ and $\lambda \sim 0.1$). Therefore
the scale factor ratio shouldn't vary too rapidly for
$t_F \leq t \leq t_{min}$ and hence will be neglected.
Then Eq. (\ref{n}) becomes
\begin{equation}
n \simeq {\lambda^{{1\over 2}(7-\alpha)}\over (1-\alpha)}
\left\{2^{(1-\alpha)} -
\left(1+{1\over k_n\lambda}\right)^{(1-\alpha)}\right\}\sigma^3.
\end{equation}
So the number density of heavy particles produced after $t_F$ is
given by $N_Q n$ and this leads to a baryon number (or lepton
number depending on the model under consideration) given by
\begin{eqnarray}
B \ (\hbox{ or } L) & \simeq & {N_Q n \over s}\epsilon \nonumber\\
& \simeq & {45\beta\over 2\pi^2 g_*}
{\lambda^{{1\over 2}(2-\alpha)}\over (1-\alpha)}
\left\{2^{(1-\alpha)} -
\left(1+{1\over k_n\lambda}\right)^{(1-\alpha)}\right\}\epsilon,
\label{BL}
\end{eqnarray}
where $\epsilon$ is the particle physics model dependent CP-violation
factor and $s\simeq {2\over 45}\pi^2 g_*\lambda^{3\over 2}\sigma^3$
is the entropy density evaluated at $T_F$.

\vskip 1cm
\leftline{\bf 3. Gauged baryon and lepton numbers}
\vskip 5mm

{}From an experimental point of view, baryon and lepton number
appear to be good symmetries, at least at the classical level.
An interesting possibility for physics beyond the SM is to
gauge the baryon and/or lepton numbers. However, no long distance
effects associated with baryon or lepton number have been
observed. So if they are gauge symmetries, then they must be
spontaneously broken.

To construct a model where $B$ and/or $L$ are gauged, the fermion
sector of the SM needs to be extended so that these symmetries
are anomaly-free. We know of two ways to do this. One way is to
add a mirror set of fermions which have the same quantum numbers
as the SM fermions but with opposite chiralities. In this case
the anomalies cancel trivially. The other way is to add  ``exotic''
generations to the SM fermions. The multiplet structure of an
exotic generation can be defined as:
\begin{eqnarray}
& \hbox{leptons}: & N_R(Y)\oplus (N-1)_L(Y-1)\oplus (N-1)'_L(Y+1)
\oplus (N-2)_R(Y); \nonumber\\
& \hbox{quarks}: & N_R\left({1\over 3}Y\right)\oplus (N-1)_L\left(
-{1\over 3}Y-1\right)
\nonumber\\
& & \oplus (N-1)'_L\left(-{1\over 3}Y+1\right)\oplus (N-2)_R\left(
-{1\over 3}Y\right),
\end{eqnarray}
where $N$ is the $N$-dimensional representation of $SU(2)_L$
and $Y$ is the hypercharge. One of the simplest anomaly-free
structures of the gauge group
\begin{equation}
G_{BL} = SU(3)_c \otimes SU(2)_L \otimes U(1)_Y
\otimes U(1)_B \otimes U(1)_L
\label{gbl}
\end{equation}
corresponds to four ordinary doublet-type and two exotic triplet-type
generations. Note that it is also possible to use the same methodology
to construct models where either $B$ or $L$ are anomaly-free only.
For further details see Ref. \cite{fjl}.

Having noted that it is possible to gauge $B$ and/or $L$, we now
consider a model with the gauge group given by Eq. (\ref{gbl})
where we will apply the results of section 2 to generate the BAU.
The fermion sector consists of the SM fermions together with a set
of either mirror fermions or exotic generations to cancel the
potential $B$ and $L$ anomalies. The usual electroweak symmetry
breaking and fermion mass generation are facilitated by the SM Higgs
doublet. In addition, two complex SM Higgs singlets, $H_1$ and $H_2$,
are introduced to spontaneously break $U(1)_B$ and $U(1)_L$ when they
acquire nonzero VEVs, $\langle \sigma e^{i\theta}\rangle$ and
$\langle \eta e^{i\phi}\rangle$, respectively. The Higgs fields
transform under the gauge group $G_{BL}$ as
\begin{equation}
H_1 \sim (1, 1, 0)(2, 0)\  \hbox{ and } \ H_2 \sim (1, 1, 0)(0, 2),
\end{equation}
where the first set of parentheses give the SM quantum numbers and the
second set gives the $B$ and $L$ numbers. From phenomenological
considerations, it is expected that the scales of $B$ and $L$ breaking
lie above the electroweak scale. Since $B$ and $L$
are spontaneously broken the formation of cosmic strings will result
and particle production from the collapse of these string loops
will follow the scenario described in section 2. In this section
we will only be interested in what happens in the quark sector
for the generation of the BAU and thus we will not consider the
lepton sector any further. So far a nonzero VEV for $H_1$ only
breaks the gauged $B$ symmetry, leaving behind a residual global
$B$ symmetry. To also break the global symmetry so that there is
baryon number violation in the quark sector, the following vectorlike
pair of fermions are introduced. Their transformation properties
under $G_{BL}$ are defined by their Yukawa interactions with $H_1$:
\begin{equation}
{\cal L}^Q_{Yuk} = \lambda_L H_1\bar{Q}_L Q_L^c
 + \lambda_R H_1\bar{Q}_R Q_R^c
 + 2D \bar{Q}_LQ_R + {\rm H.c.},
\label{Q}
\end{equation}
so that
\begin{equation}
Q_L \sim (8, 1, 0)(1, 0)\  \hbox{ and }\  Q_R \sim (8, 1, 0)(1, 0).
\end{equation}
Finally, we need to introduce one more scalar field into the model.
This is a charged colored scalar field, $\Delta$, which does not get
a nonzero VEV. Its purpose is to transfer the $B$ asymmetry produced
in the $Q_{L,R}$ sector to the ordinary quark sector. The transformation
properties of $\Delta$ are determined by the interaction term,
\begin{equation}
{\cal L}^{\Delta}_{Yuk} = \lambda_{\Delta}\bar{Q}_L\Delta d_R + {\rm H.c.},
\label{Delta}
\end{equation}
so that $\Delta \sim (\bar 3, 1, {2\over 3})({2\over 3}, 0)$, where
$d_R$ is the usual right-handed $d$-type quark field.

We will now proceed to determine the BAU from the collapse of
string loops by using the results of section 2. First we will
outline the scenario of section 2 as applied to this model. When
the Universe cools down to a temperature, $T_c \sim \sigma$, the
$U(1)_B$ symmetry is spontaneously broken and the formation of
cosmic strings is initiated. The string loops collapse rapidly
in the friction dominated epoch of the evolution of the string
network. The production of Higgs scalars, $h$ ($h$ is a real
scalar field left over from the Higgs mechanism with its mass given
by $M_h \simeq \lambda^{1\over 2}\sigma $), is assumed to result from
this collapse. This provides a source of these Higgs scalars which are
at this point overabundant and hence out of equilibrium. The scalar, $h$,
then decays rapidly into the $Q$-type particles since its decay width,
of the order of $M_{h}$, is much larger than the Universe expansion rate,
$H\sim T^2/M_{p}$, at the temperatures under consideration and thereby an
excess of $B$ can be generated. The $Q$'s in turn get converted
to $d$-type quarks via their interactions with the $\Delta$ scalars.
$B$ can be calculated from Eq. (\ref{BL}) once $\epsilon$ is
determined. Hence the remainder of this section is used to show
how $\epsilon$ can be calculated.

To determine $\epsilon$ we need to write down the interactions
of $h$ with the $Q$'s. First write $Q_{L,R}$ in terms of their
mass eigenstates. When $H_1$ develops a nonzero VEV the interaction
terms of Eq. (\ref{Q}) generates the following mass terms:
\begin{equation}
{\cal L}_{mass} = \bar\Psi_R {\cal M} \Psi_L + {\rm H.c.},
\end{equation}
where
\begin{equation}
\Psi_L = \left(\begin{array}{c}
Q_L \\ Q_R^c \end{array} \right),\quad
\Psi_R = \left(\begin{array}{c}
Q_L^c \\ Q_R \end{array}\right), \quad
{\cal M} = \left(\begin{array}{cc}
\lambda_L \sigma & D \\
D & \lambda_R \sigma \end{array}\right).
\end{equation}
Note that $\lambda_{L,R}$ can be made real by a phase redefinition
of the fields. The mass matrix, ${\cal M}$, can be diagonalized so
that
\begin{equation}
{\cal L}_{mass} = \bar\Psi_R' {\cal D} \Psi_L' + {\rm H.c.},
\end{equation}
where ${\cal D} = Diag(M_1, M_2)$ and the mass eigenstates are given by
\begin{equation}
\Psi_L' = \left(\begin{array}{c}
\zeta_{1L} \\ \zeta_{2R}^c \end{array} \right) = U^\dagger\Psi_L,\quad
\Psi_R' = \left({\begin{array}{c}
\zeta_{1L}^c \\ \zeta_{2R} \end{array}} \right) = V^\dagger\Psi_R
\end{equation}
with
\begin{equation}
U = \left(\begin{array}{cc}
\cos\theta & \sin\theta \\
-e^{i\delta}\sin\theta  & e^{i\delta}\cos\theta
\end{array}\right)\  \hbox{ and }\
V = \left(\begin{array}{cc}
\cos\theta & \sin\theta \\
-e^{-i\delta}\sin\theta  & e^{-i\delta}\cos\theta
\end{array}\right).
\end{equation}
The interactions of $h$ from Eq. (\ref{Q}) can now be written as
\begin{equation}
{\cal L}_{int} = \lambda_1 h\bar{\zeta^c}_{1L}\zeta_{1L}
+ \lambda_2 h\bar{\zeta^c}_{2R}\zeta_{2R}
+ \lambda_{12} h\bar{\zeta}_{2R}\zeta_{1L} + {\rm H.c}.,
\label{int}
\end{equation}
where
\begin{eqnarray}
\lambda_1 & = & \lambda_L\cos^2\theta
+ \lambda_R e^{i2\delta}\sin^2\theta, \nonumber \\
\lambda_2 & = & \lambda_L\sin^2\theta
+ \lambda_R e^{-i2\delta}\cos^2\theta, \nonumber \\
\lambda_{12} & = & \sin 2\theta
\left(\lambda_L - \lambda_R e^{i2\delta}\right).
\end{eqnarray}
It is important to note that $\lambda_{1,2}$ and $\lambda_{12}$
cannot be made real because the phases of the fields are now fixed
in the mass eigenstate basis. These complex couplings are necessary
for the nonzero CP-violation parameter, $\epsilon$, to be generated.
{}From Eq. (\ref{int}), it can be seen that there are three processes
which contribute to $\epsilon$. They are
(1) $h \longrightarrow \zeta_{1L}\bar\zeta_{2R}$,
(2) $h \longrightarrow \zeta_{1L}\zeta_{1L}$ and
(3) $h \longrightarrow \zeta_{2R}\zeta_{2R}$. The Feynman diagrams
for process (1) are shown in Figs. (1) and (2). The corresponding
diagrams for processes (2) and (3) are very similar. For example, for
process (1), only the diagram in Fig. 2(b) has a complex product of
couplings for the interference term between the tree and one-loop
amplitudes and hence can contribute to $\epsilon$. The relevant
amplitudes can be written as follows:
\begin{eqnarray}
{\cal A}(h \longrightarrow \zeta_{1L}\bar\zeta_{2R})
& = & \lambda_{12} \hat a_0
+ \lambda^*_{12} \lambda_1 \lambda^*_2 \hat a_1 \nonumber\\
\bar{\cal A}(h \longrightarrow \bar\zeta_{1L}\zeta_{2R})
& = & \lambda^*_{12} \hat a_0
+ \lambda_{12} \lambda^*_1 \lambda_2 \hat a_1,
\end{eqnarray}
where $\hat a_0$ and $\hat a_1$ denote the tree and one-loop contributions
respectively. Then
\begin{equation}
\vert {\cal A} \vert^2 - \vert \bar{\cal A} \vert^2
= - 4\: {\rm Im} \left(\lambda_1 \lambda^*_2 \lambda^{*2}_{12} \right)
{\rm Im} \left(\hat a_0^* \hat a_1 \right).
\end{equation}
For convenience, define
\begin{equation}
\Delta ( i \longrightarrow f ) \equiv
\Gamma ( i \longrightarrow f ) - \Gamma ( \bar i \longrightarrow \bar f ),
\end{equation}
where $i$ and $f$ represent the initial and final states of the
decay process. Then
\begin{equation}
{\Delta ( h \longrightarrow \bar\zeta_{1L}\zeta_{2R} )
\over \Gamma ( h \longrightarrow \hbox{all} ) }
\simeq
 - { 4\: {\rm Im} \left(\lambda_1 \lambda^*_2 \lambda^{*2}_{12} \right)
{\rm Im} \left(a_0^* a_1 \right) \over
\left(\vert\lambda_1\vert^2 + \vert\lambda_2\vert^2
+ 2\vert\lambda_{12}\vert^2 \right)\vert a_0 \vert^2 } \beta_{12}^3
\end{equation}
Similarly for processes (2) and (3)
\begin{equation}
{\Delta ( h \longrightarrow \zeta_{1L}\zeta_{1L} )
\over \Gamma ( h \longrightarrow \hbox{all} ) }
\simeq
 + { 2\: {\rm Im} \left(\lambda_1 \lambda^*_2 \lambda^{*2}_{12} \right)
{\rm Im} \left(a_0^* a_1 \right) \over
\left(\vert\lambda_1\vert^2 + \vert\lambda_2\vert^2
+ 2\vert\lambda_{12}\vert^2 \right)\vert a_0 \vert^2 } \beta_{11}^3
\end{equation}
and
\begin{equation}
{\Delta ( h \longrightarrow \zeta_{2R}\zeta_{2R} )
\over \Gamma ( h \longrightarrow \hbox{all} ) }
\simeq
 - { 2\: {\rm Im} \left(\lambda_1 \lambda^*_2 \lambda^{*2}_{12} \right)
{\rm Im} \left(a_0^* a_1 \right) \over
\left(\vert\lambda_1\vert^2 + \vert\lambda_2\vert^2
+ 2\vert\lambda_{12}\vert^2 \right)\vert a_0 \vert^2 } \beta_{22}^3
\end{equation}
respectively, where $a_0$ and $a_1$ are the amplitudes $\hat a_0$ and
$\hat a_1$ with the phase space factor
$\beta_{ij} = \sqrt{1 - (M_i + M_j)^2/M_h^2}$ factored out.
Note that $a_0$ and $a_1$ are not the same for each of the above
processes since $M_1 \neq M_2$ in general. However in the following,
to simplify the calculations, we will assume $M_1 \sim M_2$ and
much smaller than $M_h$. There is also a set of diagrams
where the internal Higgs line is replaced by the massive gauge
boson line but these diagrams do not contribute to $\epsilon$
since the product of couplings of the interference terms are real
due to the real gauge couplings. So far, all of the $B$-violation
is carried by $\zeta_{1,2}$ in the exotic sector but this will be
transferred to the ordinary sector via the interactions of
Eq. (\ref{Delta}), i.e.,
\begin{equation}
{\cal L}^{\Delta}_{Yuk}
= \lambda_{\Delta}\left(\cos\theta\bar{\zeta}_{1L}\Delta d_R
+ \sin\theta\bar{\zeta^c}_{2R}\Delta d_R\right) + {\rm H.c}.,
\label{Delta2}
\end{equation}
Therefore $\epsilon$ can be written as
\begin{eqnarray}
\epsilon & = & \sum_f B_{f'}
{\Delta ( h \longrightarrow f )
\over \Gamma ( h \longrightarrow \hbox{all} ) }
{\Gamma ( f \longrightarrow f' )
\over \Gamma ( f \longrightarrow \hbox{all} ) } \nonumber\\
& =  & B_d \Biggl\{
{\Delta ( h \longrightarrow \zeta_{1L}\zeta_{1L} )
\over \Gamma ( h \longrightarrow \hbox{all} ) }
{\Gamma ( \zeta_{1L} \longrightarrow \Delta^\dagger d_R )
\over \Gamma ( \zeta_{1L} \longrightarrow \hbox{all} ) } \nonumber\\
& + & {\Delta ( h \longrightarrow \bar\zeta_{2R}\bar\zeta_{2R} )
\over \Gamma ( h \longrightarrow \hbox{all} ) }
{\Gamma ( \bar\zeta_{2R} \longrightarrow \Delta^\dagger d_R )
\over \Gamma ( \bar\zeta_{2R} \longrightarrow \hbox{all} ) } \nonumber\\
& + & {1\over 2}
{\Delta ( h \longrightarrow \zeta_{1L}\bar\zeta_{2R} )
\over \Gamma ( h \longrightarrow \hbox{all} ) }
\left[ {\Gamma ( \zeta_{1L} \longrightarrow \Delta^\dagger d_R )
\over \Gamma ( \zeta_{1L} \longrightarrow \hbox{all} ) }
+ {\Gamma ( \bar\zeta_{2R} \longrightarrow \Delta^\dagger d_R )
\over \Gamma ( \bar\zeta_{2R} \longrightarrow \hbox{all} ) }
\right] \Biggr\}
\end{eqnarray}
which simplifies to
\begin{equation}
\epsilon \sim  {1\over 4\pi}{\left(M_1 - M_2\right)^2\over M_h^2}
{{\rm Im} \left(\lambda_1 \lambda^*_2 \lambda^{*2}_{12} \right)
\over \vert\lambda_1\vert^2 + \vert\lambda_2\vert^2
+ 2\vert\lambda_{12}\vert^2},
\end{equation}
where the branching ratios of the processes,
$\zeta_{1L}, \bar\zeta_{2R} \longrightarrow \Delta^\dagger d_R$
are unity, the factor
$16\pi {\rm Im} \left(a_0^* a_1 \right)/ \vert a_0 \vert^2 $ is of
order one \cite{nw} and $B_d = -{2\over 3}$ for the final state
baryon number since each process results in the decay product
$\Delta^{\dagger}\:d_R$.
To get a rough estimate of the size of the couplings required
for the observed value of the BAU, let
$\lambda_\zeta \equiv \lambda_1 \sim \lambda_2 \sim \lambda_{12}$ and
$M_{1,2} \sim {\cal O}(\lambda_\zeta \sigma )$,
so that $\epsilon \sim \lambda^4_\zeta/\left(16\pi\lambda\right)$.
Substituting this in Eq. (\ref{BL}) gives
\begin{equation}
B \sim  10^{-3} {\lambda^{-{\alpha\over 2}}
\over\left(1 - \alpha\right)}
\left\{2^{(1-\alpha)} -
\left(1+{1\over k_n\lambda}\right)^{(1-\alpha)}\right\}
\lambda^4_\zeta.
\label{B}
\end{equation}
For example, by taking $\lambda \sim 0.1$, $\alpha = 2.5$, with
$k_n \sim 10$ (see Eq. (\ref{kn})), then $\lambda_\zeta \sim 10^{-2}$
for the present baryon asymmetry which lies in the
range $(4-5.7)\times 10^{-11}$ \cite{walk}. Therefore if $M_{1,2}\lsim
10^{3}$ GeV, then the breaking scale for $B$ is expected to be less
than about $10^5$ GeV. Note that no particular limit must be set on the
$L$ breaking scale since now sphalerons violate neither $B$ nor $L$ and
therefore the interplay between the sphalerons and $L$-violating processes
is completely harmless for the generated baryon asymmetry.  Furthermore, the
dangerous $B$-violating processes like  $\Delta + \Delta\longrightarrow
d_{R} + d_{R}$ mediated by the $\zeta$ particles, whose rate is
$\sim \lambda_{\Delta}^4 T^3/m_{\zeta}^2$,
are out of equilibrium for small enough values of $\lambda_{\Delta}$.
As a result they also do not affect the $B$ produced from the rapid decays
of $\zeta \longrightarrow \Delta + d_{R}$.

\vskip 1cm
\leftline{\bf 4. The gauged ($B-L$) model}
\vskip 5mm

It is known that the SM with right-handed neutrinos has an
anomaly-free $(B-L)$ symmetry. Therefore $(B-L)$ can be gauged.
In this section we will consider such a model for generating
the BAU using the collapse of string loops mechanism of section 2.
The gauge group is
\begin{equation}
SU(3)_c \otimes SU(2)_L \otimes U(1)_Y \otimes U(1)_{B-L}
\end{equation}
To break the $(B-L)$ symmetry and to give the right-handed neutrinos
a Majorana mass, we introduce the scalar field, $\chi$, which is a
complex SM Higgs singlet with two units of $(B-L)$.
As the temperature in the early Universe drops to about
$T \sim \langle \chi \rangle$ the $(B-L)$ symmetry is spontaneously
broken and the formation of cosmic strings begins. These strings
will evolve in the friction dominated period where string loops
tend to collapse rapidly. The collapse of string loops will result
in the production of Higgs particles associated with $\chi$. These
Higgs particles will then in turn decay into right-handed neutrinos
providing a source of lepton number. By using the results of section 2,
$L$ will be given by Eq. (\ref{BL}). The $\epsilon$ parameter can be
calculated from the decays of the right-handed neutrino, {\it i.e.}
\begin{eqnarray}
\nu_{R_{i}} & \rightarrow & F_{L_{j}} + \bar\Phi \nonumber \\
\nu_{R_{i}} & \rightarrow & \bar F_{L_{j}} + \Phi,
\end{eqnarray}
such that
\begin{equation}
\epsilon = \sum_i \epsilon_i
\end{equation}
where $\epsilon_{i}$ is the difference between particle-antiparticle
branching ratios given by \cite{FY}
\begin{equation}
\epsilon_{i}=\frac{1}{2\pi\left(h h^{\dagger}\right)_{ii}}
\sum_{j}\left({\rm Im}\left[\left(h h^{\dagger}\right)_{ij}
\right]^2\right) f\left(m_{\nu_{R_{j}}}^2/m_{\nu_{R_{i}}}^2\right),
\end{equation}
with
\begin{equation}
f(x)=\sqrt{x}\left[1-\left(1+x\right){\rm ln}\left(\frac{1+x}{x}\right)
\right].
\end{equation}
The $h$'s denote the Yukawa couplings between right-handed
and left-handed neutrinos through the SM Higgs doublet $\Phi$ and are
assumed to be complex to give a source of CP violation.

The lepton number produced by right-handed neutrino
decays is converted to a nonzero baryon number via the $(B-L)$
conserving sphaleron processes. Therefore the BAU is given by
\begin{eqnarray}
B & = & \kappa\: L \nonumber \\
  & \sim & \kappa\: {45\beta\over 2\pi^2 g_*}
{\lambda^{{1\over 2}(2-\alpha)}\over (1-\alpha)}
\left\{2^{(1-\alpha)} -
\left(1+{1\over k_n\lambda}\right)^{(1-\alpha)}\right\}\epsilon,
\end{eqnarray}
where $\kappa$ is a numerical factor of ${\cal O}(1)$ and
can be easily calculated from Ref. \cite{har}.
For $\lambda \sim 10^{-2}$ and $\alpha =2.5$ gives a baryon
asymmetry of about $B \sim 10^{-1} \epsilon$. For the present
baryon asymmetry to be of the order of $5\times 10^{-11}$,
$\epsilon$ has to be as large as $10^{-10}$.

We also need to check that the initial baryon asymmetry generated
by the collapsing string loops and the subsequent decays of right-handed
neutrinos is not erased by a combination of other lepton violating
interactions and sphaleron processes \cite{FY2}. The lowest dimension
$L$-violating operator in the effective low energy theory is given by
\begin{equation}
{\cal L}_{\Delta L = 2}
= {m_{\nu_L}\over v^2}F_L F_L \Phi\Phi + {\rm H.c.},
\end{equation}
where $m_{\nu_L}$ is the mass of the left-handed neutrino
and $v$ is the electroweak breaking scale. The interaction
rate of these $\Delta L = 2$ processes is
$\Gamma_{\Delta L = 2}
\simeq m_{\nu_L}^2 T^3/ \left(\pi^3 v^4 \right)$. For the
survival of the pre-existing asymmetry we require this interaction
rate to be less than the expansion rate of the Universe, {\it i.e.},
$H \sim T^2/ M_p$, where $T$ is given by $T_{min}$ in Eq. (11).
This results in a bound on the $(B-L)$ breaking scale,
\begin{equation}
\langle\chi \rangle
\gsim \frac{\gamma^4}{\pi^3 h^2}k_{n}^{-1/3}\lambda^{1/2}\: M_{p},
\end{equation}
where $h$ is the Yukawa coupling for the Dirac mass term and $\gamma$ is
the one for the Majorana mass. So for natural values of the couplings
involved, the bound on $\langle \chi \rangle$ can easily be made as
low as $10^{9}$ GeV.

\vskip 1cm
\leftline{\bf 5. Conclusions}
\vskip 5mm

In the present paper we have shown that models, in which $B$ and $L$ are
gauged symmetries, are interesting not only from the particle physics
perspective but also for cosmology -- the generation of the baryon asymmetry
in the early Universe. Indeed, such models provide a natural implementation
of the out-of-equilibrium condition through the collapse of string loops
formed during the phase transition at intermediate scales where $B$ and/or
$L$ are spontaneously broken. Also, differently from what happens in
ref. \cite{bdh}, these loops decay during the friction dominated epoch
of string evolution. We would also like to stress that, in the case in which
both $B$ and $L$ are gauged, the sphaleron processes do not violate $B+L$
so that there is no wash out of any pre-existing asymmetry and no particular
limit must be imposed on the scale of $L$ breaking (other than that from
experiment). Finally, in our scenario, the electroweak phase transition need
not be first order as required in most non-GUT scenarios for the generation
of the BAU \cite{rev2}.

\vskip 1cm
\centerline{\bf Acknowledgements}
\vskip 1mm
\noindent
It is a pleasure to express our
gratitude to A. Masiero for stimulating and enlightening
discussions. This work was supported in part by a grant from the DOE.

%% This section generates the figures. If you don't want the figures
%% either remove the following section or insert \end{document} before
%% this section

%fig

%%%%%%%%%%%%%%%%%%%%%% figures %%%%%%%%%%%%%%%%%%%%%%%%%%%%%%%%%%%%%%%%

\vskip 2cm
\textwidth7in
\oddsidemargin-.2in
\def\lappeq{\mathrel{\rlap{\raise.5ex\hbox{$<$}}
{\lower.5ex\hbox{$\sim$}}}}
\baselineskip=21pt

\input FEYNMAN
\bigphotons
%Figure 1
\begin{picture}(40000,10000)
\drawline\scalar[\E\REG](10000,0)[4]
\global\advance\pmidy by 500
\put(\pmidx,\pmidy){$h$}
\drawline\fermion[\SE\REG](\pbackx,\pbacky)[8000]
\global\advance\pmidy by -1250
\global\advance\pmidx by -2400
\put(\pmidx,\pmidy){$\bar{\zeta}_{2R}$}
\drawline\fermion[\NE\REG](\pfrontx,\pfronty)[8000]
\global\advance\pmidy by 700
\global\advance\pmidx by -1900
\put(\pmidx,\pmidy){$\zeta_{1L}$}
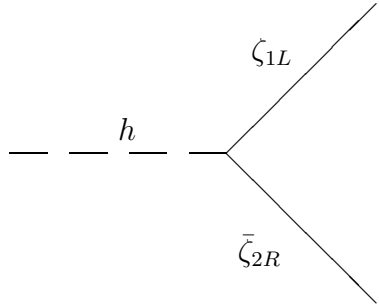
\put(0,-20000){ Figure 1:  Tree level diagram for the decay
$h\rightarrow\zeta_{1L}\bar{\zeta}_{2R}$.}
\end{picture}

%Figure 2(a)-(d)
\begin{picture}(20000,50000)
\put(0,52000){(a) }
\drawline\scalar[\E\REG](2000,52000)[4]
\global\advance\pmidy by 500
\put(\pmidx,\pmidy){$h$}
\drawline\fermion[\SE\REG](\pbackx,\pbacky)[7400]
\global\advance\pmidy by -1250
\global\advance\pmidx by -2400
\put(\pmidx,\pmidy){$\zeta_{1L}$}
\drawline\fermion[\NE\REG](\pfrontx,\pfronty)[7400]
\global\advance\pmidy by 700
\global\advance\pmidx by -1900
\put(\pmidx,\pmidy){$\zeta_{2R}$}
\drawline\fermion[\E\REG](\pbackx,\pbacky)[7400]
\global\advance\pmidy by 500
\put(\pmidx,\pmidy){$\zeta_{1L}$}
%\drawline\fermion[\S\REG](\pfrontx,\pfronty)[11330]
\drawline\scalar[\S\REG](\pfrontx,\pfronty)[5]
\global\advance\pmidx by -200
%\put(\pmidx,\pmidy){$\times$}
\global\advance\pmidx by 1500
\put(\pmidx,\pmidy){$h$}
\drawline\fermion[\E\REG](\pbackx,\pbacky)[7400]
\global\advance\pmidy by 500
\put(\pmidx,\pmidy){$\bar{\zeta}_{2R}$}

\put(0,37000){(b) }
\drawline\scalar[\E\REG](2000,37000)[4]
\global\advance\pmidy by 500
\put(\pmidx,\pmidy){$h$}
\drawline\fermion[\SE\REG](\pbackx,\pbacky)[7400]
\global\advance\pmidy by -1250
\global\advance\pmidx by -2400
\put(\pmidx,\pmidy){$\zeta_{2R}$}
\drawline\fermion[\NE\REG](\pfrontx,\pfronty)[7400]
\global\advance\pmidy by 700
\global\advance\pmidx by -1900
\put(\pmidx,\pmidy){$\zeta_{1L}$}
\drawline\fermion[\E\REG](\pbackx,\pbacky)[7400]
\global\advance\pmidy by 500
\put(\pmidx,\pmidy){$\zeta_{1L}$}
%\drawline\fermion[\S\REG](\pfrontx,\pfronty)[11330]
\drawline\scalar[\S\REG](\pfrontx,\pfronty)[5]
\global\advance\pmidx by -200
%\put(\pmidx,\pmidy){$\times$}
\global\advance\pmidx by 1500
\put(\pmidx,\pmidy){$h$}
\drawline\fermion[\E\REG](\pbackx,\pbacky)[7400]
\global\advance\pmidy by 500
\put(\pmidx,\pmidy){$\bar{\zeta}_{2R}$}

\put(0,22000){(c) }
\drawline\scalar[\E\REG](2000,22000)[4]
\global\advance\pmidy by 500
\put(\pmidx,\pmidy){$h$}
\drawline\fermion[\SE\REG](\pbackx,\pbacky)[7400]
\global\advance\pmidy by -1250
\global\advance\pmidx by -2400
\put(\pmidx,\pmidy){$\zeta_{1L}$}
\drawline\fermion[\NE\REG](\pfrontx,\pfronty)[7400]
\global\advance\pmidy by 700
\global\advance\pmidx by -1900
\put(\pmidx,\pmidy){$\zeta_{1L}$}
\drawline\fermion[\E\REG](\pbackx,\pbacky)[7400]
\global\advance\pmidy by 500
\put(\pmidx,\pmidy){$\zeta_{1L}$}
%\drawline\fermion[\S\REG](\pfrontx,\pfronty)[11330]
\drawline\scalar[\S\REG](\pfrontx,\pfronty)[5]
\global\advance\pmidx by -200
%\put(\pmidx,\pmidy){$\times$}
\global\advance\pmidx by 1500
\put(\pmidx,\pmidy){$h$}
\drawline\fermion[\E\REG](\pbackx,\pbacky)[7400]
\global\advance\pmidy by 500
\put(\pmidx,\pmidy){$\bar{\zeta}_{2R}$}

\put(0,7000){(d) }
\drawline\scalar[\E\REG](2000,7000)[4]
\global\advance\pmidy by 500
\put(\pmidx,\pmidy){$h$}
\drawline\fermion[\SE\REG](\pbackx,\pbacky)[7400]
\global\advance\pmidy by -1250
\global\advance\pmidx by -2400
\put(\pmidx,\pmidy){$\zeta_{2R}$}
\drawline\fermion[\NE\REG](\pfrontx,\pfronty)[7400]
\global\advance\pmidy by 700
\global\advance\pmidx by -1900
\put(\pmidx,\pmidy){$\zeta_{2R}$}
\drawline\fermion[\E\REG](\pbackx,\pbacky)[7400]
\global\advance\pmidy by 500
\put(\pmidx,\pmidy){$\zeta_{1L}$}
%\drawline\fermion[\S\REG](\pfrontx,\pfronty)[11330]
\drawline\scalar[\S\REG](\pfrontx,\pfronty)[5]
\global\advance\pmidx by -200
%\put(\pmidx,\pmidy){$\times$}
\global\advance\pmidx by 1500
\put(\pmidx,\pmidy){$h$}
\drawline\fermion[\E\REG](\pbackx,\pbacky)[7400]
\global\advance\pmidy by 500
\put(\pmidx,\pmidy){$\bar{\zeta}_{2R}$}
\global\advance\pmidy by -3500
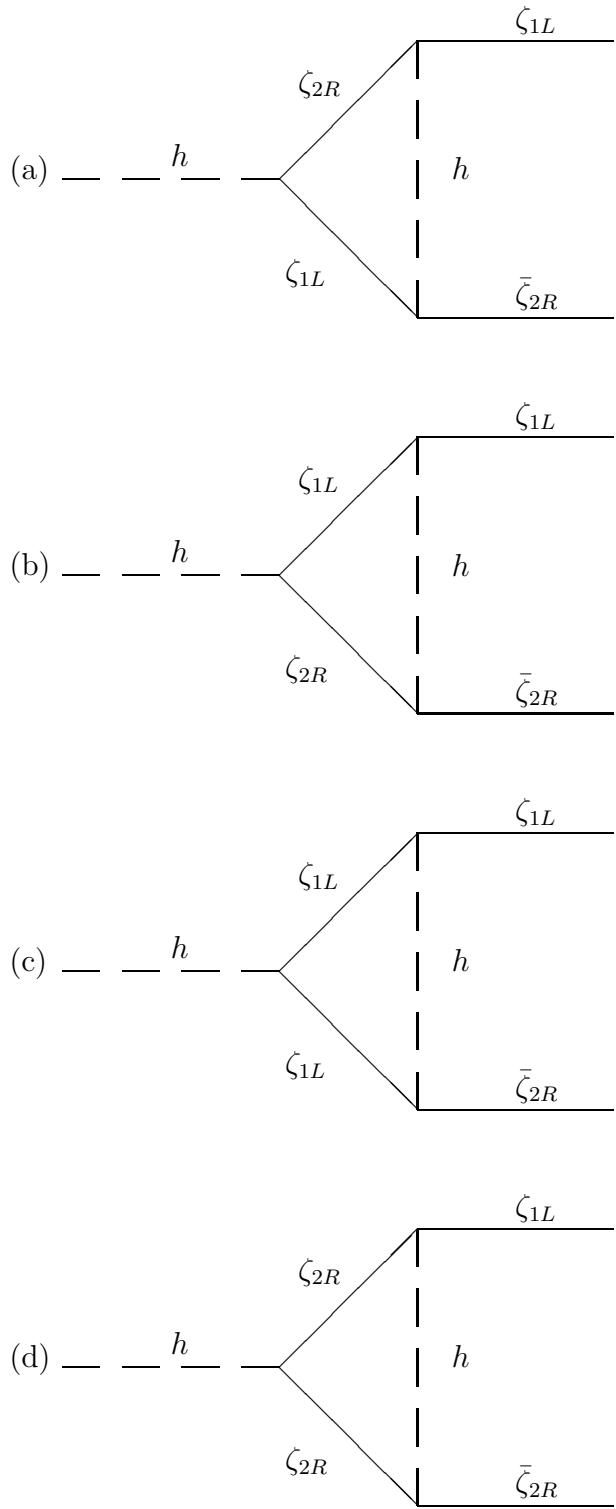
\put(0,\pmidy){Figure 2: One-loop diagrams contributing to CP violation
in the $h\rightarrow\zeta_{1L}\bar{\zeta}_{2R}$ decay;}
\global\advance\pmidy by -1500
\put(0,\pmidy){only diagram (b) gives a nonvanishing contribution.}
\end{picture}

%%%%%%%%%%%%%%%%%%%%%% end of figures %%%%%%%%%%%%%%%%%%%%%%%%%%%%%%%%%%%%%%

\end{document}